\documentclass[showpacs,twocolumn,prl]{revtex4}
\usepackage{graphicx}
\usepackage{epsfig}
\usepackage{amssymb}

\def\beq{\begin{equation}}
\def\eeq{\end{equation}}
\def\beqa{\begin{eqnarray}}
\def\eeqa{\end{eqnarray}}

\def\e{\epsilon}

\def\half{{\ss 1\over 2}}

\def\D{\Delta}

\def\e{\epsilon}
\def\cH{{\mathcal H}}
\def\cL{{\mathcal L}}
\def\ss{\scriptstyle}
\def\hcon{\mathrm{H.c.}}
\def\Tr{\mathrm{Tr}}
\def\L{\mathrm{L}}
\def\R{\mathrm{R}}
\def\tip{\mathrm{tip}}
\def\r{{\bf r}}

\def\d{\mathrm{d}}

\def\ss{\scriptstyle}
\def\al{\alpha}

\def\etal{{\sl et al.}}

\def\cdag{c^{\dagger}}

\renewcommand{\sec}[1]{\vskip 0.truecm \noindent \emph{#1}. -- }
\newcommand{\ket}[1]{| #1 \rangle}
\newcommand{\bra}[1]{\langle #1 |}

\begin{document}

\title{Reconstructing Fourier's law from disorder in quantum wires}
\author{Y. Dubi and M. Di Ventra}
\affiliation{Department of Physics, University of California San Diego, La Jolla, California 92093-0319, USA}
\begin{abstract}
The theory of open quantum systems is used to study the local temperature and heat currents in metallic nanowires connected to
leads at different temperatures. We show that for ballistic wires the local temperature is almost uniform along the wire and
Fourier's law is invalid. By gradually increasing disorder, a uniform temperature gradient ensues inside the wire and the thermal
current linearly relates to this local temperature gradient, in agreement with Fourier's law. Finally, we demonstrate that while
disorder is responsible for the onset of Fourier's law, the non-equilibrium energy distribution function is determined solely by
the heat baths.
\end{abstract}
\maketitle The search for a microscopic derivation of Fourier's law
\cite{Fourier}, or even a microscopic demonstration of it, is still
a major theoretical challenge \cite{Bonetto,Buchanan}. In recent
years, the problem seems more relevant than ever due to both the
continuous miniaturization of electronic devices and the need for
alternative energy sources. Both these trends require better
understanding of fundamental processes of energy transport in
nanoscale systems. Following, there have been many theoretical
attempts to derive Fourier's law in various systems, both classical
and quantum
\cite{Michel,Lepri,Garrido,Michel1,Bricmont,Gaspard,Roy}. In the
quantum regime, attention has been focused mainly on small
spin-chains \cite{Michel,Michel1,Mejia} or quantum harmonic
oscillators \cite{Gaul}, where it was demonstrated that Fourier's
law holds for chaotic systems, but vanishes when transport turns
ballistic. This is in agreement with the behavior of classical
systems, where it has long been postulated that chaos leads to
Fourier's law \cite{Michel}.

However, little is known on the energy transport in nanoscale electronic quantum systems. One of the reasons for this is that in
order to demonstrate the validity - or violation - of Fourier's law one needs to (i) define a local temperature out of
equilibrium and show that it develops a uniform gradient, (ii) evaluate the local heat current, and (iii) show proportionality
between these two quantities. The first task is especially difficult since temperature is a global equilibrium property, and it
is not clear if a "local temperature" can be defined at all when the system is out of equilibrium \cite{Nagi,Hartmann}. For this
reason, recent studies of the origin of Fourier's law either use a phenomenological definition of local temperature (as an
expectation value of a local energy operator \cite{Michel,Wang}) or assume that a temperature gradient is already present
\cite{Wu}. An alternative route is to study the energy diffusion in closed systems (i.e. without thermal baths)
\cite{Steinigeweg}.

An additional reason that renders calculation of energy transport in electronic systems a formidable task is the fact that the
size of the Hilbert space scales exponentially with the number of electrons, making numerical calculations very demanding. This
is why previous numerical calculations on heat transport in quantum systems usually refer to very small systems, typically of the
order of ten spins (see, however, Ref.~\cite{Prosen}).

Here we report a calculation of energy transport in electronic quantum wires that overcomes both the above issues. It is based on
solving the quantum master equation for non-interacting electrons in the presence of dissipative baths (held at different
temperatures) in the Markov approximation. We use a recently suggested method \cite{Pershin,Dubi} to map the many-electron
problem to a single-particle system, which allows calculations for systems an order of magnitude larger than in previous studies,
and enables us to define a local temperature operationally, that is one directly accessible experimentally, even out of
equilibrium.

\sec{Model} The system consists of a linear chain bonded to small
leads, which are connected to thermal baths held at different
temperatures (upper panel of Fig. 1).~\cite{Pershin,Dubi} A similar
set-up was used for spin-chains in, e.g., Ref. \cite{Mejia2}. The
Hamiltonian of the system is given by $
\cH=\cH_L+\cH_R+\cH_d+\cH_{c} $, where $\cH_{L,R,d}=\sum_{i \in
\L,\R,\d} \e_i \cdag_i c_i   -t \sum_{\langle i,j \rangle \in
\L,\R,\d} \left(\cdag_i c_j + h.c.\right)$ are the tight-binding
Hamiltonians of the left lead ($\L$), right lead ($\R$) and wire
($\d$, of length $L_d$), respectively ($t$ is the hopping integral,
which serves as the energy scale hereafter). $\cH_{c}=
(g_L\cdag_{L}c_{d,0}+g_R\cdag_{R}c_{d,L_d}+h.c. )$ describes the
coupling between the left (right) lead to the wire. $\cdag_{L(R)}$
are creation operators for an electron at the point of contact
between the left (right) lead and the wire, and $c_{d,0}$
($c_{d,L_d}$) destroys an electron at the left-most (right-most)
sites of the wire. The on-site energies $\e_i$ are randomly drawn
from a uniform distribution $U[-W/2,W/2]$, with $W$ being the
disorder strength. The lattice constant is taken to be $a=1$, and we
consider here spinless electrons.

\begin{figure}
\vskip 0.5truecm
\includegraphics[width=7truecm]{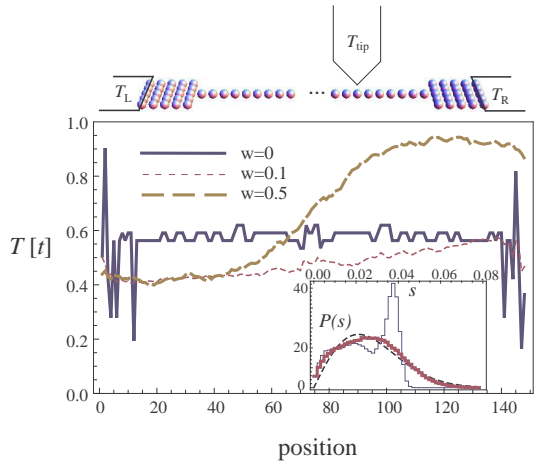}
\caption{(color online) Local temperature as a function of position along the wire. The local temperature is calculated for three
different values of disorder, $W=0$ (solid), $0.1$ (small dashing), and $0.5$ (large dashing). For clean wires ($W =0$) the
temperature is uniform, and a uniform temperature gradient builds up as the disorder increases. Upper panel: geometry of the
model system. The solid lines at the edges correspond to the contact area of the thermal baths. Inset: energy level spacing
distribution for $W=0.1$ (thin line) and $W=0.5$ (thick line). For weak disorder the distribution shows a structure reminiscent
of the clean system. For strong disorder the distribution resembles the Wigner-Dyson distribution (dashed line), marking the
onset of chaos. }\label{fig1}
\end{figure}

The quantity of interest, from which the required information (such as local temperature, density and heat current) may be
extracted, is the single-particle density matrix, defined by $\rho=\sum_{kk'}\rho_{kk'}\ket{k}\bra{k'}$ , where $\rho_{kk'}=\Tr
(c^\dagger_k c_{k'} \hat{\rho_{{\mathrm MB}}}) $ , $\hat{\rho_{{\text MB}}}$ is the full many-body density-matrix, and $\ket{k} $
are the single-particle states. In Ref.~\cite{Pershin,Dubi} it was shown that $\rho$ obeys a master-equation of the Lindblad form
\cite{Lindblad} (setting $\hbar=1$) \beq \dot{\rho}=-i[\cH,\rho]+\cL_{\L}[\rho]+\cL_{\R}[\rho] \equiv \hat{\cL}[\rho ],\eeq where
$\cL_{\L(\R)}[\rho]$ are super-operators acting on the density matrix, describing the left (right) thermal baths, held at
temperature $T_{\L(\R)}$, and in contact with the left (right)-most side of the leads (solid lines in the upper panel of Fig. 1).
The super-operators are defined in the Lindblad form \cite{Lindblad,Van Kampen} via V-operators, $\cL [\rho] = \sum_{k,k'} \left(
-\half \{ V^\dagger_{kk'} V_{kk'}, \rho \} +V_{kk'} \rho V^\dagger_{kk'} \right) $, with $\{\cdot,\cdot\}$ being the
anti-commutator. The $V$-operators are generalized to account for the different baths, and are given by \cite{Dubi,Mejia2} $
V^{(L,R)}_{kk'}=\sqrt{\gamma^{(L,R)}_{kk'}f^{(L,R)}_D(\e_k)} \ket{k} \bra{k'} $, where $f^{(L,R)}_D(\e_k)=1 /\left( \exp \left
(\frac{\e_k-\mu}{k_BT_{L,R}}\right)+1 \right)$ are the Fermi distributions of the left and right leads, with $\mu$ the chemical
potential. The coefficients $ \gamma^{(L,R)}_{kk'}= \left| \sum_{\r_i\in S_{L,R}} \psi_k(\r_i)\,\gamma_0\,\psi^*
_{k'}(\r_i)\right|$ (where $\psi_{k}(r) $ are the single-particle wave functions) describe the overlap between the
single-particle states $\ket{k}$ and $\ket{k'}$ over the region of contact $S_{L(R)}$ between the left (right) baths and the
corresponding junction leads, shown by the solid lines in the upper panel of Fig.~\ref{fig1}. $\gamma_0$ describes the strength
of electron-phonon (bath) interaction. The form above can be derived from first principles by tracing out the bath degrees of
freedom, with the latter formed by a dense spectrum of boson excitations (e.g., phonons), which interact {\em locally} with
electrons at the edges of the system.

\sec{Local temperature} The non-equilibrium steady state of the
system is given by the solution of the equation $\hat{\cL}[\rho ]=0
$. In order to calculate the local temperature, we attach a third
thermal bath (described by an additional term $\cL_{\tip}(r)[\rho] $
in the master equation) which is connected to a given position r of
the wire and serves as an external probe (upper panel of Fig. 1).
The resulting equation for the steady state with inclusion of this
probe now reads $(\hat{\cL}+\cL_{\tip}(r))[\rho ]=0 $ . The
temperature $T_\tip$ of the additional operator is changed, and the
resulting local density, $n(r)=\sum_k \rho_{kk}|\psi_{k}(r)|^2  $
\cite{note1}, is compared to that obtained without the additional
operator. We then define the local temperature as that $T_\tip$ for
which there is minimal change in the local density.~\cite{Dubi}
Physically, this corresponds to placing a local temperature probe in
close proximity to the wire \cite{Cahill}. When the probe has the
same local temperature of the wire, there is no heat flow between
them and thus the local properties of the wire are unchanged.

Figure \ref{fig1} shows the local temperature profile along a chain of length $L_d=150$. The numerical parameters are: lead
dimensions $L_x=L_y=3$, temperature of the left and right leads is $T_\L=0.1$ and $T_\R=1$ respectively, and electron number
$n_E=56$ (corresponding to one third filling).  The disorder strength is $W=0$ (solid line), $W=0.1$ (small dashing) and $W=0.5$
(large dashing) and averaged over 500 realizations of disorder (all calculations in this report were performed over a wide range
of parameters, yielding similar conclusions). One can see several features from Figure 1. The most prominent feature is the fact
that for clean ($W=0$) and weakly disordered ($W=0.1$) wires the temperature hardly changes along the wire. This is clear
evidence that Fourier's law is in fact violated under these conditions. Instead, a uniform temperature gradient at the center of
the wire is recovered for large disorder ($W=0.5$). In addition, from Fig. 1 one can see an asymmetry between low and high
temperatures - the local temperature of the wire never reaches that of the colder bath (with $T_\L=0.1$), but does reach the
hotter bath temperature (with $T_\R=1$). This asymmetry can be attributed to the structure of the Fermi function, which gives
different weights to high and low temperatures. Also notice the thermal length at the edges of the wire, where the local
temperature is roughly constant.

In order to determine a relation between the appearance of a uniform
temperature gradient and the onset of chaos \cite{Michel}, the inset
of Fig. 1 shows the distribution function $P(s)$ for the
level-spacing $s$ of neighboring single-particle energy levels, for
$W=0.1$ (thin line) and $W=0.5$ (thick line). The dashed line is the
Wigner-Dyson distribution, $P_{\mathrm{WD}}(s)=s \exp (-\alpha s^2)$
(where $\al$ is some constant), which is conjectured to correspond
to the onset of quantum chaos \cite{Guhr}. As seen, the distribution
for weak disorder shows features reminiscent of the ordered system
and is very different from the Wigner-Dyson distribution.
Accordingly Fourier's law is invalid. On the other hand, for strong
disorder we see both agreement with the Wigner-Dyson distribution
and a uniform temperature gradient at the center of the wire,
suggesting that indeed these have the same physical origin.

\sec{Heat current} The next step in validating Fourier's law comes from calculating the local heat current $j(r)$, and evaluating
the thermal conductivity $\kappa$ such that $j(r)=-\kappa (r) \nabla T(r) $ . It is clear that for a uniform temperature profile
(as that of a clean wire), a finite heat current results in a divergent $\kappa$.  On the other hand, for the system presented
here one expects that the heat current is always finite, since energy is always injected and extracted from the system. It is
thus natural to define a global thermal conductivity, $K=-\frac{L_d \bar{j}}{T_\R-T_\L} $, where $\bar{j} $  is the heat current
averaged over the whole wire. Following this definition, it is evident that $K=\bar{\kappa}$ (where $\bar{\kappa}$ is the average
over the wire of the local thermal conductivity, defined via Fourier's law) only when Fourier's law is valid. This definition
thus gives us a mean to point to the onset of Fourier's law.

To make this argument substantial one must calculate the local heat current. Since there is some ambiguity in defining the local
heat current, we follow Ref.~\cite{Wu,Wu1} and define the heat current operator via a time-derivative of the local energy
operator $\hat{\cH}_i$, i.e. $\frac{\d \hat{\cH_i}}{\d t}=\hat{j}_{i-1}-\hat{j}_i $ (in units of $t^2$, taking $\hbar$=1). The
local energy operator in the wire is defined as $\cH_i=\e_i \ket{i}\bra{i}-\half t (\ket{i}\bra{i+1}+\ket{i}\bra{i-1}+\hcon)$.
The time-derivative of a general operator $\hat{O}$ in the Lindblad formalism is given by \cite{Van Kampen} $ \frac{\d
\hat{O}}{\d t}=i [\cH,\hat{O}]+\sum_{kk'}\left(-\half \{ V^\dagger_{kk'} V_{kk'}, \hat{O} \} +V_{kk'} \hat{O} V^\dagger_{kk'}
\right)$. From this relation it is straightforward to find $\hat{j}_i $ and to calculate its expectation value, $j_i$. Fig. 2(a)
shows $j_i$ as a function of position along the wire (same parameters as in Fig. 1), for different values of disorder $W=0,
0.1,..., 0.5$. Note that in the central region of the wire the heat current is uniform, and increases at a distance of the order
of the thermal length of the leads (solid lines in the $W=0.5$ curve are guides to the eye), due to the proximity of the heat
baths. The inset of Fig. 2(b) shows the averaged heat current (over the whole wire) as a function of temperature difference $\D
T=T_\R-T_\L$, for different values of disorder. From the linear regime one can extract the global $K$, shown by empty circles in
Fig. 2(b) as a function of disorder. On the other hand, one can extract the averaged local $\kappa$ by using the local heat
current (Fig. 2(a)) and the local temperature (from Fig. 1), calculated and averaged over sites close to the center of the wire.
The values of the local $\kappa$ are shown as filled circles in Fig. 2(b), and exhibit a divergence for clean wires. Only at
large disorder, where Fourier's law is valid, do the two definitions of thermal conductivity coincide. In fact, we can take this
as criterion for the onset of Fourier's law, which microscopically corresponds to the value of disorder $W$ of order $t/2$,
namely when the strength of the disorder is comparable to half of the kinetic energy parameter.

\begin{figure}
\vskip 0.5truecm
\includegraphics[width=7truecm]{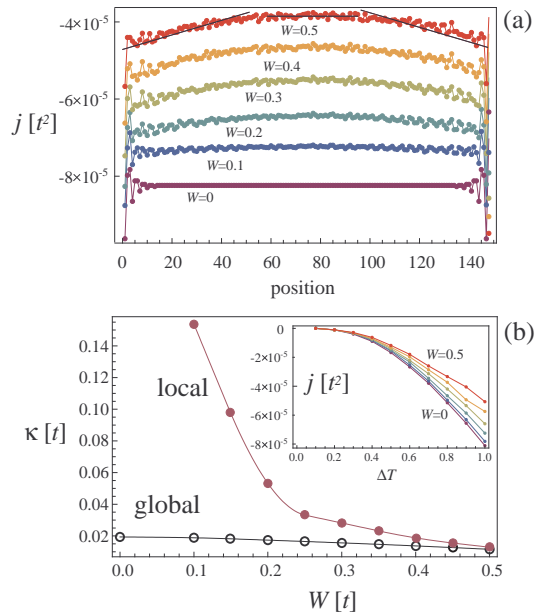}
\caption{(color online) Local heat current and thermal conductivity. (a) Local heat current j as a function of position for
different disorder strengths, $W=0, 0.1…0.5$. The current is largest close to the leads, where the contact with the heat baths is
located. The solid line on W=0.5 is a guide to the eye. (b) Local (solid circles) and global (empty circles) thermal conductivity
as a function of disorder. While the global thermal conductivity hardly changes with disorder, the local thermal conductivity
shows strong disorder dependence, and diverges in the clean limit due to the vanishing temperature gradient. The two tend to the
same value only with increasing disorder, when Fourier's law becomes valid. Inset: averaged heat current as a function of
temperature difference $\D T$ for different disorder strengths. From the linear regime of these curves the global thermal
conductivity was evaluated.  }\label{fig2}
\end{figure}

In order to further determine the role of disorder, we study the local and global electron energy distribution function, both of
which can be measured experimentally \cite{Pothier}. In terms of the density matrix, the global distribution function is simply
$f(E_k)=\rho_{kk} $, and the local distribution function is given by $f_\mathrm{loc}(E_k,r)=|\psi_k(r)|^2 \rho_{kk}$. Fig. 3
shows the distribution function for the clean wire ($W=0$, squares) and disordered wire ($W=0.5$, circles, same parameters as in
Fig. 1) as a function of energy. As seen, there is hardly a difference between the two functions. The solid line is the curve
$f(E)=\half (f_\L (E)+f_\R (E))$ , where $ f_{\L (\R)}(E)$ is the Fermi distribution of the left (right) lead, with the
corresponding temperature. The excellent agreement between the numerical curves and the averaged density suggests that one cannot
consider the wire as a system with an effective temperature $T_{\mathrm{eff}}=\half (T_\L+T_\R)$ ,but rather it is the
distribution function itself that is averaged, and that disorder does not affect the distribution function.

In the inset of Fig. 3 we plot the local distribution function for the clean and disordered wires at the left edge of the wire
(thick line) and at the center of the wire (thin line). Again, the curves for clean and disordered wires are very similar. We
note that in our geometry, in contrast to mesoscopic wires \cite{Pothier}, one cannot attribute to the local distribution a
simple position-dependence, even in the presence of disorder.

\begin{figure}
\vskip 0.5truecm
\includegraphics[width=7truecm]{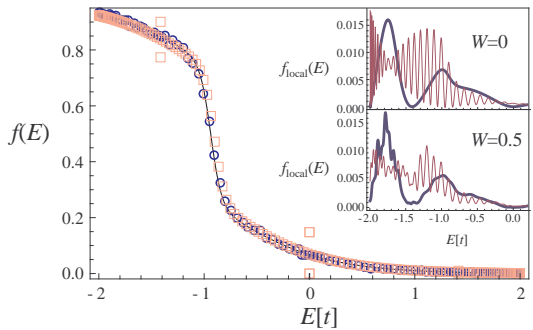}
\caption{(color online) Global electron distribution functions with and without disorder. The distribution function is evaluated
for a clean wire (squares) and disordered wire, $W=0.5$ (circles). The thick line corresponds to an average distribution
$f(E)=\half (f_\L (E)+f_\R (E))$ , where $ f_{\L (\R)}(E)$ is the Fermi distribution of the left (right) lead, with the
corresponding temperature. Inset: the local electron distribution function at the edge of the wire (thick line) and the center of
the wire (thin line) for a clean system (upper panel) and a disordered system (lower panel). The resemblance between the two
cases demonstrates that the distribution function is determined not by the local electronic structure, as dictated by the
Hamiltonian, but rather by the boundary conditions, i.e. the thermal baths. }\label{fig3}
\end{figure}

These findings show that the energy distribution function is only determined by the boundary conditions, namely those provided by
the bath operators, and not by the local structure (i.e. the Hamiltonian). On the other hand, nature of the heat transport is
determined solely by the microscopic character of the Hamiltonian states. This implies that Fourier's law cannot be validated
from measuring the local distribution function.

One possible way to verify Fourier's law is to study the local temperature. The local temperature of electronic systems can be
measured experimentally \cite{Cahill}. Relevant experimental systems for which our predictions may be tested can be, e.g., carbon
nanotubes \cite{Yu}, quantum point contacts \cite{van Houten}, atomic-size metallic wires \cite{Ludoph} or silicon nanowires
\cite{Hochbaum,Boukai}, where measurements of thermal conductance and thermo-power have already been demonstrated. A different
route is to measure the length-dependence of the thermal conductivity, as was recently applied to nanotubes \cite{Chang}

 Finally we note that the above theory did not include the effects of electron interactions. While for some quasi-one-dimensional
 systems electron interactions may not be very important, it is nevertheless of interest to find out whether electron interactions
  alone suffice for the validity of Fourier's law, even in clean wires. However, the above theory is limited to non-interacting systems, and to
resolve this issue one must employ a more elaborated method that can encompass both interactions and environments \cite{Di
Ventra}. Such studies are currently underway.

This work is supported by DOE under grant DE-FG02-05ER46204.


\begin{thebibliography}{}
\bibitem{note1}
Note that the off-diagonal elements of the density matrix vanish at the steady state, see Ref.~\cite{Pershin}.
\end{thebibliography}
\end{document}